\documentclass[12pt]{iopart}

  \usepackage{iopams}
  \usepackage{setstack}
  \usepackage{graphicx}
  \usepackage{color}
  \usepackage{amssymb}
  \usepackage{amsbsy}

\begin{document}

\title{Anomalous diffusion in viscosity landscapes}

\author{M. Burgis$^1$, V. Schaller$^{1,2}$, M. Gl\"assl$^{1}$, B. Kaiser$^{1}$, 
  W. K\"ohler$^{3}$, A. Krekhov$^{1}$, W. Zimmermann$^{1}$}

\address{$^{1}$ Theoretische Physik I, Universit\"at Bayreuth,  95440 Bayreuth, Germany\\
$^{2}$ Biophysik E27, Technische Universit\"at M\"unchen,  85748 Garching, Germany\\
$^{3}$ Experimentalphysik IV, Universit\"at Bayreuth,  95440 Bayreuth, Germany}

\ead{walter.zimmermann@uni-bayreuth.de}

\date{\today}

%%%%%%%%%%%%%%%%%%%%%%%%%%%%%%%%%%%%%%%%%%%%%
\begin{abstract}
%%%%%%%%%%%%%%%%%%%%%%%%%%%%%%%%%%%%%%%%%%%%%
Anomalous diffusion is predicted for Brownian particles in
inhomogeneous viscosity landscapes by means of  scaling arguments,
which are substantiated through numerical simulations. Analytical solutions of the related Fokker-Planck equation
in limiting cases confirm our results.
For an ensemble of particles starting at 
a spatial minimum (maximum) of the viscous damping 
we find subdiffusive (superdiffusive) motion.
Superdiffusion occurs also for a monotonically varying viscosity profile.
We suggest different substances for related experimental investigations.
\end{abstract}
%%%%%%%%%%%%%%%%%%%%%%%%%%%%%%%%%%%%%%%%%%%%%

%%%%%%%%%%%%%%%%%%%%%%%%%%%%%%%%%%%%%%%%%%%%%
\section{Introduction} \label{sec: intro}
%%%%%%%%%%%%%%%%%%%%%%%%%%%%%%%%%%%%%%%%%%%%%

Brownian particle dynamics plays a key role for 
many transport processes in various disciplines.
Even though a century has passed since the publication
of Einstein's seminal work  on {\it normal diffusion} \cite{Einstein:1905.1}, 
the field of diffusion processes still attracts enormous interest
\cite{Dhont:96,Risken:84,Haenggi:2005.1,Gardiner:09}.
The essential signature of normal diffusion is the 
linear temporal growth of the mean square displacement 
of Brownian particles,
\begin{eqnarray}
\label{msq}
 \langle x^2(t) \rangle \propto t^\alpha \,,
\end{eqnarray}
with  $\alpha=1$, whereas for $\alpha \neq 1$
the notion {\it anomalous diffusion} has been introduced 
\cite{Klafter:2005.1,Klafter:2005.2,SokolovRad:2009,Klafter:2000.1,Montroll:1975.1}. 
Being not only of fundamental but also of great 
technological relevance,
the dynamical processes underlying anomalous diffusion phenomena 
have started to be investigated during recent decades 
in systems as diverse as 
collective ordering phenomena \cite{Chate:2008,Ramaswamy:2007}, 
particle diffusion in mesoscopic systems \cite{HeinrichD:2008.1_PRL_101,Hoefling:2008},
or social networks \cite{Brockmann:2006.1}. 
At the same time, progress in single particle manipulation and detection 
on the $\mu$m and sub-$\mu$m scale fostered investigations on random particle motions 
\cite{Kosztolowicz:2005.1_PRL_94,
Cox:2006.1_PRL_96,Fradin:2005.1,weitz:2004.1,Schwab:2008.1}.

Diffusion processes with an exponent $0< \alpha<1 $
in (\ref{msq}) are called {\it subdiffusive}. 
Several examples of such dynamic behavior  are known for 
systems with a short inhomogeneity length scale,
 such as in  amorphous semiconductors  \cite{Montroll:1975.1},
  groundwater motion \cite{Denz:2004.1}, 
diffusion in gels \cite{Kosztolowicz:2005.1_PRL_94}, 
and in several biological systems such as
in the bacterial cytoplasm \cite{Cox:2006.1_PRL_96}, 
during protein diffusion through cell membranes \cite{Fradin:2005.1}, or
as a possible method to detect the microstructures in 
actin filament networks \cite{weitz:2004.1}. Besides
subdiffusive motion, heterogeneous parameter landscapes are considered
to be  important also in generalized reaction-diffusion systems 
and for bifurcations in pattern forming systems in general  \cite{Cox:2006.1_PRL_96,Baer:2009.1,Zimmermann:1993.2,anomalpatt}. 
Conversely, {\it superdiffusive} processes are random motions 
with an exponent $\alpha >1$, which are 
found in rather different areas of natural and social sciences.
Superdiffusion tends to occur in active and driven systems where 
random steps are interrupted by intermediate and nearly deterministic
motion, such as for particles in (turbulent) random velocity fields 
\cite{Procaccia:1984.1,Isichenko:1992.1}, 
during cell migration \cite{Schwab:2008.1}, 
for active transport in cells \cite{HeinrichD:2008.1_PRL_101}, or
during the spread of diseases \cite{Brockmann:2006.1}.

Another class of Brownian motion that attracted remarkable interest recently 
is described by a nonlinear Langevin equation with a
velocity dependent viscous damping coefficient 
which is used as a model for active and relativistic motion
\cite{Viscek:1995.1,Schweitzer:1998.1,Lindner:2007.1,Dunkel:2005.1} 
as well as in the context of ratchet models \cite{Jayannavar:2002.1}.

Here, we investigate the effects of inhomogeneous viscosities on 
the one-dimensional dynamics  of Brownian  particles, 
considering viscosity variations, which are slow
on the scale of the particles' random steps.
In this case, the interplay of a diffusing particle
with its inhomogeneous surrounding leads to intermediate regimes of anomalous diffusion, similar to the intermediate Rouse regime of  diffusion
of  polymer segments \cite{deGennes:79}.
Near a viscosity minimum we find subdiffusion, whereas 
 superdiffusive behavior is found near a viscosity maximum 
and for monotonously
varying viscosity profiles. In all three cases 
the particles' probability distribution is found to be non-Gaussian.

Our analysis is based on a nonlinear Langevin equation and 
the appendant Fokker-Planck equation, which are presented in
section~\ref{sec: model} together with the spatially varying model viscosities
considered in this work.
In section \ref{scaleresults} we calculate scaling formulas for the mean square displacement 
and analytical solutions of the Fokker-Planck equation for several limiting cases.
These results are compared with numerical 
simulations in section~\ref{nummeth}.
Finally, we comment on possible experimental realizations
within our summary in section~\ref{conclusion}.

%%%%%%%%%%%%%%%%%%%%%%%%%%%%%%%%%%%%%%%%%%%%%
\section{Model} \label{sec: model}
%%%%%%%%%%%%%%%%%%%%%%%%%%%%%%%%%%%%%%%%%%%%%

%%%%%%%%%%%%%%%%%%%%%%%%%%%%%%%%%%%
\subsection{Langevin and Fokker-Planck equation} \label{LE_FPE}
%%%%%%%%%%%%%%%%%%%%%%%%%%%%%%%%%%%

The one-dimensional Langevin equation for the velocity $v(t)$ of a particle
of mass $m$ and radius $a$ immersed in a medium with a spatially varying viscosity
landscape $\eta(x)$ is given by 
\begin{equation} \label{basicequation}
\dot{v}(t)=-\gamma(x) v + g(x) \xi(t)\,,
\end{equation}
with $ \gamma(x)=6 \pi a \eta(x)/m $ as the associated damping.
 The right hand side 
includes white noise  $\xi(t)$ characterized by
\begin{eqnarray}
 \langle\xi(t)\rangle=0\, \qquad \mbox{and} \qquad \langle\xi(t)\xi(t')\rangle =\delta(t-t') \,.
\end{eqnarray}
For the noise strength $g(x)$ we assume a
local fluctuation-dissipation relation \cite{Klimontovich:1994}
\begin{eqnarray} \label{fdt}
 g^2(x)=\frac{ 2 \gamma(x) k_B T}{m}\,
\end{eqnarray} 
with $k_B$ being the Boltzmann constant and $T$ denoting the constant temperature. 
In the overdamped limit, i.~e. for  times longer than the
characteristic time $\gamma^{-1}$, 
a Langevin equation for the particle's position can be derived
from (\ref{basicequation}) by using the method of adiabatic 
elimination \cite{Risken:84,Gardiner:09,SanMiguel:1982.1}.
In Ito's interpretation it takes the form
\begin{eqnarray} 
\dot{x}(t) 
&=\frac{k_{B}T}{m}\frac{\partial}{\partial x}\frac{1}{\gamma(x)}+\sqrt{\frac{2k_{B}T}{m\gamma(x)}}\ \xi(t)\,.
\label{xprop}
\end{eqnarray}
The corresponding Fokker-Planck equation 
for the probability density of a particle $P(x,t)$ can be deduced
straightforwardly from (\ref{xprop}), 
\begin{eqnarray} 
\frac{\partial P(x,t)}{\partial t}
&=\frac{\partial}{\partial x}\left[D(x)\frac{\partial P(x,t)}{\partial x}\right]\,,
\label{FPG}
\end{eqnarray}
with a spatially varying
diffusion coefficient $D(x) = k_{B}T/(m \gamma(x))$.
The Brownian dynamics of an ensemble of test particles
is investigated on the basis of (\ref{basicequation}), (\ref{xprop}) and (\ref{FPG})
with their initial positions at $x(t=0)=0$.

%%%%%%%%%%%%%%%%%%%%%%%%%%%%%%%%%%
\subsection{Model viscosities \label{modvisc}}
%%%%%%%%%%%%%%%%%%%%%%%%%%%%%%%%%%%
Spatially varying viscosities may be realized
with organic gradient materials, where for instance the
composition or the degree of polymerization changes in 
space \cite{Amis:2002.1,SchubertUS:2003.1}. 
In photorheological fluids, as another example, the
viscosity can be tuned  by a  spatially varying illumination intensity \cite{Raghavan:2007.1}.
A further class with the possibility of inhomogeneous viscosities are 
binary-fluid mixtures. Here spatial variations
of the concentration of the two constituents may be 
driven by temperature modulations via the Soret effect \cite{deGroot:84}. 
If both constituents have
sufficiently different viscosities the thermally induced concentration variations
 are accompanied by spatial viscosity changes. 
For materials with a strong Soret effect, quantified by the 
Soret coefficient $S_T$, small 
temperature gradients are sufficient to
 generate large concentration and therefore large viscosity gradients, so that
the direct effects of temperature gradients can be neglected.
 Suchlike experimentally favorable large values of $S_T$ can be achieved
by shifting the mean temperature of the binary fluid
in the one-phase region close to the critical temperature of the mixture,
where a transition to the two-phase region takes place 
\cite{KoehlerW:2004.3,KoehlerW:2007.3,KoeKreZim:2010,aizpiri:1992.1}.

For the materials mentioned above, one can imagine 
a number of spatially varying viscosities.
A simple example of a viscosity of experimental relevance, 
being asymmetric with respect to the particles' initial position at $x=0$,
is 
\begin{equation} \label{visctanh}
\eta(x)=\eta_0 +\Delta\eta\tanh (x/L) \qquad \mbox{with} \qquad \eta_0>|\Delta\eta|>0\,,
\end{equation}
which may be approximated for $|x|\ll L$ by
\begin{eqnarray} \label{visca}
 \eta(x)= \eta_0 + \tilde \eta_1 x \qquad \mbox{with} \qquad \tilde \eta_1=\Delta\eta/L\,.
\end{eqnarray}
Other generic viscosities are symmetric with a minimum or maximum
at the  particles' starting point. 
As a representative model of these viscosities,
showing a smooth change from the value $\eta_0$ at the extremum  
to the bulk value $\eta_\infty$,
 we choose
\begin{eqnarray} \label{viscb}
\eta(x) & =& \eta_\infty +  
\displaystyle\frac{\Delta \eta}{1+\left|x/L\right|^\kappa}\,,
\end{eqnarray}
with $\kappa>0$, the viscosity contrast 
$\Delta \eta= \eta_0 - \eta_\infty$, and the  characteristic length $L$. 
$\Delta \eta<0$ corresponds to a viscosity with a global minimum and 
$\Delta \eta >0$ to a maximum at $x=0$. 

Equation (\ref{viscb}) covers several limiting cases, 
which can be identified by
expressing $\eta(x)$ in (\ref{viscb}) 
for $|x| < L$  by a power series
$\eta=\eta_{\infty}+\Delta\eta\sum_{n=0}^{\infty}(-1)^{n}\left|\frac{x}{L}\right|^{n\kappa}$ around $x=0$
and for $ |x|>L$ by an asymptotic series
$\eta=\eta_{\infty}+\Delta\eta\sum_{n=0}^{\infty}(-1)^{n}\left|\frac{x}{L}\right|^{-(n+1)\kappa}$.
In distinct parameter ranges each series
can reasonably be approximated by the leading contribution, which
is feasible for the analytical considerations in section~\ref{scaleresults}.

For a pronounced minimum, i.~e. $ \eta_0 \ll \eta_\infty$, two
approximations for $\eta(x)$ are useful: 
$\eta\simeq \eta_{0}~+\eta_{\infty}|x/L|^{\kappa}$ for $|x|\ll L$ respectively 
$\eta\simeq \eta_{\infty}\left(1-|x/L|^{-\kappa}\right)$ for $|x|\gg L$.
The viscosity reaches its constant bulk value $\eta_\infty$
 for large values of $|x|$ and for $\eta_0/\eta_\infty \ll |x/L|^\kappa \ll 1$
 the latter approximation evaluates to the power law
\begin{equation}
\label{powermin}
\eta(x)\propto|x|^\kappa\,
\end{equation}
that covers the $x$ dependence of $\eta(x)$ rather well.
The validity of this power-law can be extended over a wider range with decreasing ratio   $\eta_0/\eta_\infty$.
For a pronounced  maximum, i.~e. $\eta_0 \gg \eta_\infty$, we approximate $\eta(x)$
for $|x|\ll L$ as $\eta \simeq \eta_{0}\left(1-|x/L|^{\kappa}\right)$ and 
for $|x|\gg L$ as $\eta \simeq \eta_{\infty}+\eta_{0}|x/L|^{-\kappa}$.
Accordingly for $1 \ll |x/L|^{\kappa} \ll \eta_0/\eta_\infty$
the viscosity can again be simplified to a power law
\begin{equation}
\label{powermax}
\eta (x) \propto |x|^{-\kappa} \,,
\end{equation}
while for  $ |x/L|^{\kappa} \gg \eta_0/\eta_\infty$ 
a nearly constant viscosity with $ \eta(x) \approx \eta_\infty$ results. 
The range, where the viscosity can reasonably 
be approximated by a power-law, increases with rising values of the ratio $\eta_0/\eta_\infty$.

%%%%%%%%%%%%%%%%%%%%%%%%%%%%%%%%%%%%%%%%%%%%%
\section{Scaling arguments and analytical solutions}
\label{scaleresults}
%%%%%%%%%%%%%%%%%%%%%%%%%%%%%%%%%%%%%%%%%%%%%

For symmetric viscosity profiles, $\eta(|x|)$, we present 
a scaling analysis for the power law of the mean square displacement 
$\langle x^2 \rangle \propto t^\alpha$
and in limiting cases exact solutions of the Fokker-Planck equation (\ref{FPG}). 
For asymmetric profiles we use a perturbation series to
gain the time evolution of the first moments.

%%%%%%%%%%%%%%%%%%%%%%%
\subsection{Scaling arguments for symmetric viscosity profiles}
\label{scaleresults_sym}
%%%%%%%%%%%%%%%%%%%%%%

The mean square displacement of a particle $\langle x^2(t)\rangle$
is commonly used to characterize its random motion. 
For a constant viscosity, the analytical solution of equation (\ref{basicequation}) 
takes the well-known form \cite{Dhont:96}:
 $\langle x^2(t) \rangle = \frac{2 k_B T}{m \gamma^2} 
\left( e^{-\gamma t}+\gamma t - 1 \right)$.
Beyond 
a short period of ballistic motion  in the regime of
normal diffusion, $t\gg\gamma^{-1}$,
 the mean square displacement 
increases linearly in time: 
\begin{eqnarray}
 \langle x^2(t) \rangle = \frac{2 k_B T}{m \gamma} t 
= \frac{k_B T}{3 \pi a} \frac{t}{\eta} =Q \frac{t}{\eta}\,.
\label{msqclassic}
\end{eqnarray}
In order to achieve progress by analytical calculations
also for spatially varying and 
symmetric viscosity profiles $\eta(x)=\eta(|x|)$, we use (\ref{msqclassic}) with
the replacement $\eta \to \eta(|x|)$  and
simultaneously the
substitution $|x| \to \sqrt{ \langle x^2(t) \rangle}$. The resulting expression,
\begin{eqnarray} \label{msq_approx}
 \langle x^2(t) \rangle = Q ~\frac{t}{\eta\!\left(\sqrt{\langle x^2(t)\rangle}\right)}\,,
\end{eqnarray}
is further analyzed for different regimes 
of the model viscosity (\ref{viscb}).

Within a very short time interval,
where $\langle x^2\rangle\approx 0$ holds,
the model viscosity $\eta\!\left(\sqrt{\langle x^2(t)\rangle}\right)$ 
given by (\ref{viscb})
can be approximated by $\eta_0$ ({\it regime I}) and in the
long time regime, i. e. for $\langle x^2\rangle\to\infty$, 
by $\eta_\infty$ ({\it regime III}).
In both cases one obtains with equation (\ref{msq_approx}) \emph{normal diffusion}:
%
% \begin{equation}
% \langle x^{2}(t)\rangle=\frac{Q}{\eta_{0}}~t  ~~(\mbox{{\it regime I}})
% ~~\quad\mathrm{and} ~~\quad\langle x^{2}(t)\rangle=\frac{Q}{\eta_{\infty}}~t~~
% (\mbox{{\it regime III}})\,.
% \end{equation}
\begin{equation}
\langle x^{2}(t)\rangle=\frac{Q}{\eta_{0}}~t \qquad\mathrm{and} \qquad\langle x^{2}(t)\rangle=\frac{Q}{\eta_{\infty}}~t \,.
\end{equation}
Between these two regimes, one finds {\it regime II} with a power-law behavior 
$\eta(x)= \tilde\eta ~|x|^\beta$ respectively $D(x)=\tilde D|x|^{-\beta}$
(see also (\ref{FPG}),  (\ref{powermin}) and (\ref{powermax})) for 
 $\beta=\kappa>0$ near a viscosity minimum and $\beta=-\kappa<0$ near a maximum. 
This yields together with (\ref{msq_approx})
% the expression
%
\begin{equation} \label{msd_normal}
 \langle x^2(t) \rangle = 
Q ~\frac{t}{\tilde\eta~ \langle x^2(t)\rangle^{\beta/2}  }\,,
\end{equation}
which is the basis of the prediction of \emph{anomalous diffusion}
in regime II:
\begin{equation} \label{msd_anomal}
 \langle x^{2}(t)\rangle=\left(\frac{Q}{\tilde{\eta}}\right)^\alpha~t^{\alpha}\qquad\mathrm{with}\qquad\alpha=\frac{2}{2+\beta}\,.
\end{equation}
Near a viscosity minimum, $\beta>0$, one has \emph{subdiffusion}
with $\alpha <1$ 
and near a viscosity maximum, $\beta <0$, \emph{superdiffusion}
with $\alpha>1$. For $\beta=-2$ the scaling formula (\ref{msd_anomal}) 
 breaks down and the overdamped Langevin equation (\ref{xprop}) evaluates to $\dot{x} = x~[2\tilde{D}+ (2\tilde{D})^{1/2} \,\xi(t)]\,$, which describes 
the so-called geometric Brownian motion. In this case
the power law in equation (\ref{msq}) is replaced by an
exponential time dependence of the second moment $\langle x^{2}(t)\rangle\propto e^{6\tilde{D}t}$ \cite{Oksendal:2003}.

During the crossovers between the three  regimes 
the exponent $\alpha$ of the mean square displacement, 
$\langle x^2(t)\rangle \propto t^\alpha $, varies as a function of time
from 1 in regime I to $2/(2+\beta)$ in regime II and back to 1 in
regime III. The location of regime II follows by translating 
the respective spatial ranges described in section \ref{modvisc}
onto time via (\ref{msq_approx}): 
\begin{eqnarray} \label{charact_times_1}
 \frac{2L^{2}\eta_{0}}{Q}\left(\frac{\eta_{0}}{\eta_{\infty}}\right)^{2/\kappa} \ll t\ll\frac{L^{2}\eta_{\infty}}{2Q}
    \qquad \mathrm{(subdiffusion)} \,, \\
 \frac{L^{2}\eta_{0}}{2Q} \ll t\ll\frac{2L^{2}\eta_{\infty}}{Q}\left(\frac{\eta_{0}}{\eta_{\infty}}\right)^{2/\kappa}
   \qquad \mathrm{(superdiffusion)} \,. 
\label{charact_times_2}
\end{eqnarray}
Therefore, the experimental accessibility of the anomalous regime II 
is enhanced, when the 
viscosity contrast $|\Delta \eta|=|\eta_0-\eta_\infty|$ is enlarged.

%%%%%%%%%%%%%%%%%%%%%%
%%%%%%%% Fokker-Planck
%%%%%%%%%%%%%%%%%%%%%%
\subsection{Fokker-Planck equation: Solutions for symmetric and
asymmetric viscosity profiles } \label{FPE_analytic_solution}

For a power law approximation $\eta(x)\propto|x|^{\beta}$ 
of symmetric viscosity profiles in 
regime II the diffusion coefficient 
 is given  by $D(x)=\tilde{D}|x|^{-\beta}$.
In this case, the Fokker-Planck equation (\ref{FPG}) with the initial condition
$P(x,0)=\delta(x)$ can be solved analytically for $\beta >-2$
 by using the ansatz
\begin{equation}
\label{FPanalytic}
	P(x,t) = \frac{k}{t^{1/(2+\beta)}}\exp\!\left(-\frac{|x|^{2+\beta}}{t\tilde{D}(2+\beta)^{2}}\right)\,,
\end{equation}
with $k=\frac{2+\beta}{2}\,[\tilde{D}(2+\beta)^{2}]^{-1/(2+\beta)}\,\Gamma^{-1}(\frac{1}{2+\beta})$ 
being a normalization constant. With this expression the second moment 
can be evaluated exactly to
\begin{equation} \label{FPanalytic_msd}
	\langle x^{2}\rangle\, = \frac{\Gamma(\frac{3}{2+\beta})}{\Gamma(\frac{1}{2+\beta})}\,\left[\tilde{D} (2+\beta)^{2} \, t\right]^{2/(2+\beta)}  \, \propto\, t^{2/(2+\beta)}\,,
\end{equation}
which follows the identical power-law in time as 
in (\ref{msd_anomal}) and supports the validity of the
 assumptions made for the scaling argument in section \ref{scaleresults_sym}.
The odd moments vanish because of the $\pm x$ symmetry and
the even moments deviate for $\beta\neq 0$ from the case of normal diffusion.
As an example we consider the kurtosis $K$,
which is a quantity for the distribution's peakedness and
defined as 4th central moment divided by the 2nd central moment \cite{SokolovRad:2009}.
For (\ref{FPanalytic}) it takes the form $K=\Gamma\!\left(\frac{1}{2+\beta}\right) \Gamma\!\left(\frac{5}{2+\beta}\right) \Gamma^{-2}\!\left(\frac{3}{2+\beta}\right)$,
indicating a normal distribution ($K=3$, i.~e. mesokurtic) for $\beta=0$,
 a distribution with a high sharp peak ($K>3$, i.~e. leptokurtic) for $-2<\beta<0$, and 
 a flat-topped distribution ($K<3$, i.~e. platykurtic) for $\beta>0$.

%%%%%%%%%%%%%%%%%%%%%%
%%%%%%%% Asymmetric
%%%%%%%%%%%%%%%%%%%%%%

To gain insight in the diffusion for a 
 asymmetric linear viscosity profile as in equation 
 (\ref{visca}), respectively for the corresponding diffusion
coefficient, $D(x)=\tilde{D}(\eta_{0}+\tilde{\eta}_1x)^{-1}$
with $  \tilde D = \frac{k_BT}{6\pi a}$, we derive the 
equation for the $n$-th moment,
\begin{equation}
\frac{\partial}{\partial t}\langle x^{n}\rangle=n\left\langle
\frac{\partial}{\partial x}D(x)x^{n-1}\right\rangle \,,
\label{fpe_moments}
\end{equation} 
as a function of $D(x)$ by a double integration by parts
of the Fokker-Planck equation (\ref{FPG}).
In order to proceed we expand on the right hand side of equation 
(\ref{fpe_moments}) the term inside the brackets 
with respect to powers of $x$.
The resulting system of 
coupled differential equations for the moments $\langle x^{n}\rangle$
is solved in the  limit $\tilde{\eta}_1 x/\eta_0 \ll 1$ 
by a power series ansatz in $\tilde \eta_1$. 
This leads together with the initial conditions $\langle x^{n}\rangle(0)=0$
to the following approximations for the first three moments:
\numparts
\label{moments_asym}
\begin{eqnarray}
%\label{moments_asym}
\langle
x\rangle&=&-\frac{\tilde{\eta}_1\tilde{D}}{\eta_{0}^{2}}~t-\frac{4\tilde{\eta}_1^{3}\tilde{D}^{2}}{\eta_{0}^{5}}~t^{2}+\mathcal{O}(\tilde{\eta}_1^{5})
\,, \label{mean_asym} \\
\langle
x^{2}\rangle&=&~\frac{2\tilde{D}}{\eta_{0}}~t+\frac{8\tilde{\eta}_1^{2}\tilde{D}^{2}}{\eta_{0}^{4}}~t^{2}+\frac{280\tilde{\eta}_1^{4}\tilde{D}^{3}}{3\eta_{0}^{7}}~t^{3}+\mathcal{O}(\tilde{\eta}_1^{6})
\,, \label{msd_asym}  \\
\langle x^3 \rangle& =& -\frac{12 \tilde \eta_1 \tilde D^2 }{\eta_0^3}~t^2
-\frac{140 \tilde \eta_1^3 \tilde D^3 }{\eta_0^6} ~t^3 +{\mathcal O}(\tilde \eta_1^5)\,.
\label{third_asym}
\end{eqnarray}
\endnumparts
According to (\ref{msd_asym}) anomalous diffusion can be expected to occur for
linearly varying viscosity profiles, too.  Further, the onset of anomalous diffusion
is governed by the ratio $\tilde \eta_1/\eta_0$: The larger this ratio the sooner
the transition to anomalous diffusion takes place. Asymmetric viscosity profiles
lead to a   mean particle drift towards the region with lower viscosity 
as indicated by  (\ref{mean_asym}).
It is also a remarkable property of each of the three moments 
that the sign of the leading contributions does not alternate.

%%%%%%%%%%%%%%%%%%%%%%%%%%%%%%%%%%%%
\section{Numerical results} \label{nummeth}
%%%%%%%%%%%%%%%%%%%%%%%%%%%%%%%%%%%%

The results of the previous section are substantiated in the following
by comparisons with numerical simulations of the Langevin equation~(\ref{basicequation})
and the Fokker-Planck equation~(\ref{FPG}) for different spatially varying viscosities.

\begin{figure}[htb]
  \begin{center}
	\includegraphics[width=0.49\columnwidth,angle=-0]{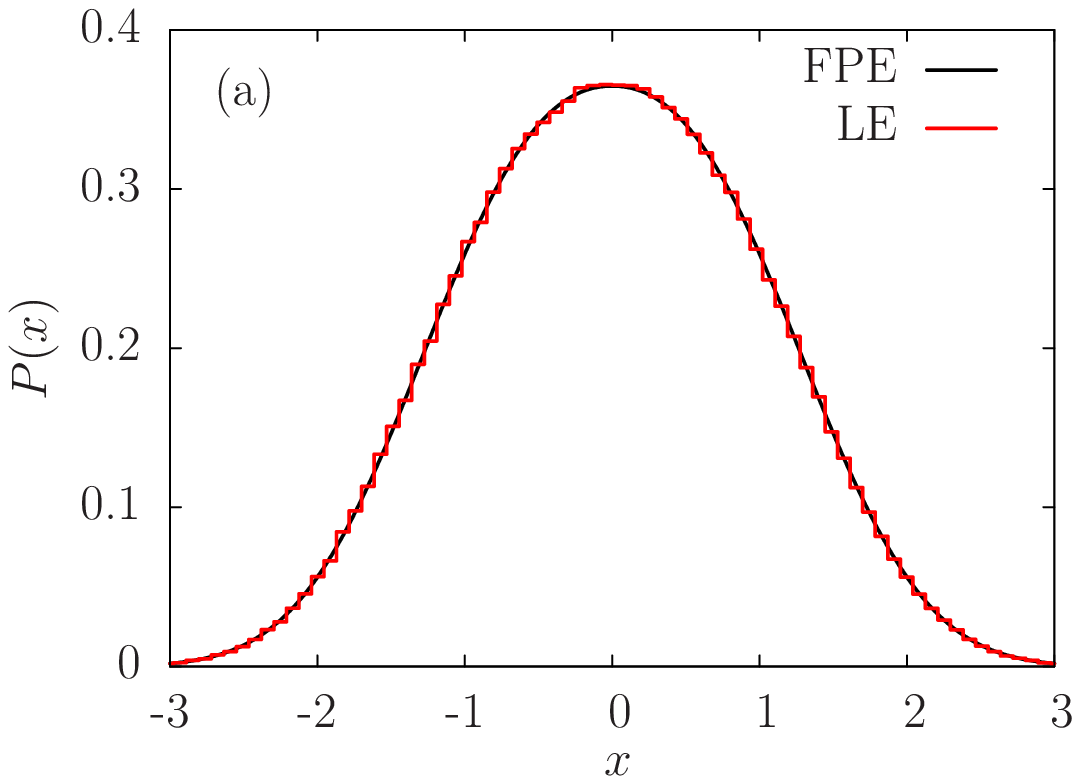}
	\includegraphics[width=0.49\columnwidth,angle=-0]{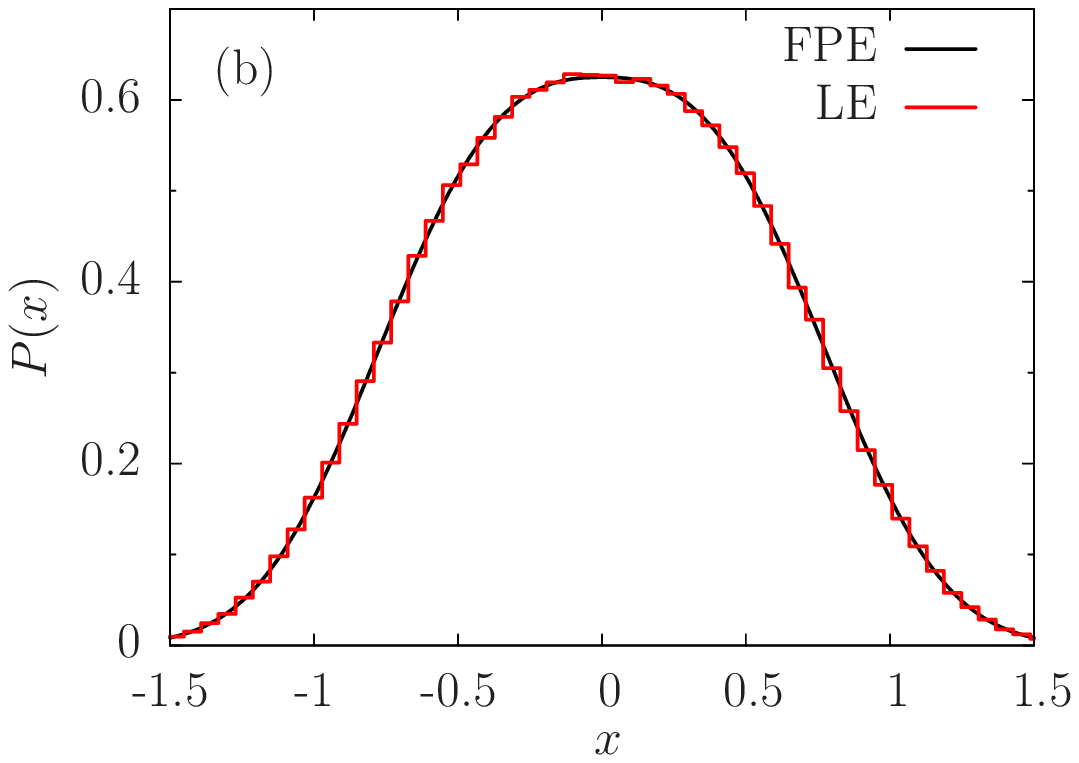}
 \end{center}
  \caption{\label{distributeF} 
   The probability density function $P(x)$ at $t=10$ is given
for $\eta(x)=\eta_0 +\tilde \eta |x|$ with $\eta_0=10$ and 
$\tilde \eta=10$ in (a) and $\tilde  \eta=100$ in (b). 
The solid line (FPE) is obtained as a numerical solution 
of (\ref{FPG}) with a narrow initial distribution 
and the step function (LE) by integrating 
(\ref{basicequation}) for  $10^6$ independent trajectories
with the starting point at $x=0$.
}
\end{figure}
The Fokker-Planck equation~(\ref{FPG}) with a spatially varying damping is 
derived from (\ref{basicequation}) under the assumption
$t\gg\gamma^{-1}$ \cite{SanMiguel:1982.1}.
For the symmetric viscosity 
 $\eta(x)=\eta_0+\tilde \eta |x|$ with the minimal value $\eta_0=10$ 
and two different slopes $\tilde \eta=10$ and $\tilde \eta=100$, 
the distribution $P(x)$ is determined in figure~\ref{distributeF} at
the time $t=10\gg \gamma_\mathrm{min}^{-1}=0.1$
by solving either the Fokker-Planck equation (\ref{FPG}) (black solid curve)
 or by integrating the Langevin equation (\ref{basicequation}) 
for an ensemble of $10^6$ particles (red step function).
The inequality $t\gg\gamma^{-1}$ is not always fulfilled
as good as in figure~\ref{distributeF}, but the solutions of
(\ref{FPG}) are also
for shorter times and stronger variations of $\eta(x)$ 
often in good agreement with the solutions of (\ref{basicequation}).

\begin{figure}[htb]
\begin{center}
\includegraphics[width=0.49\columnwidth,angle=-0]{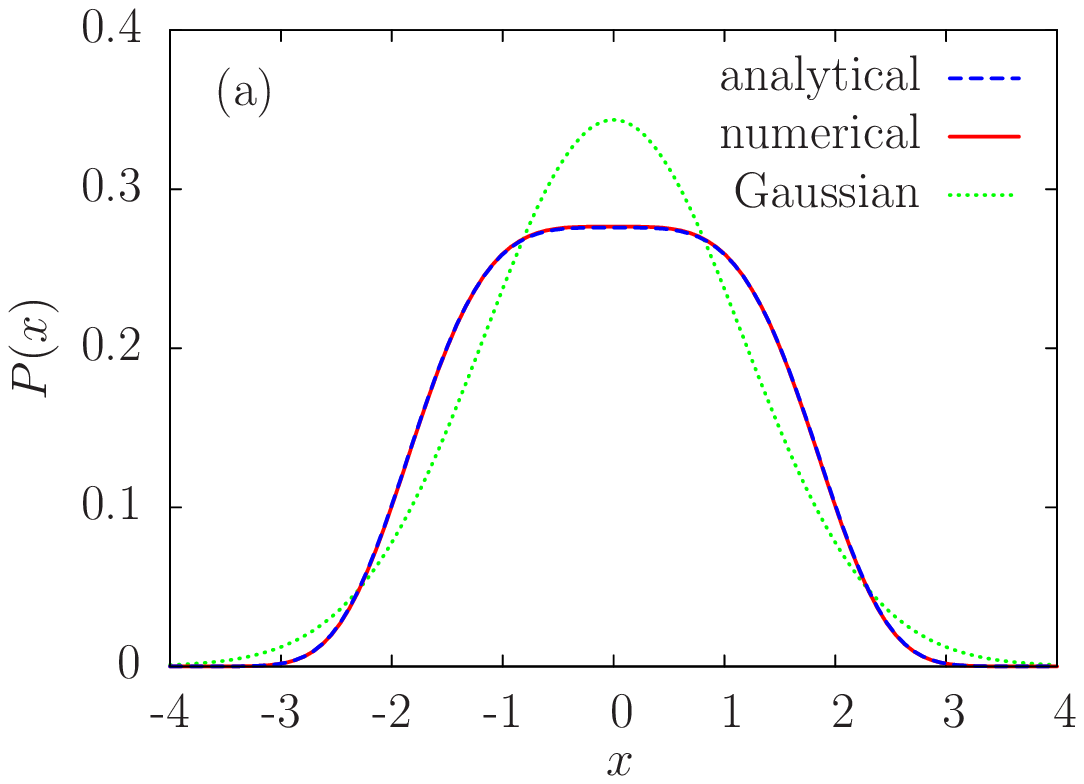}
\includegraphics[width=0.49\columnwidth,angle=-0]{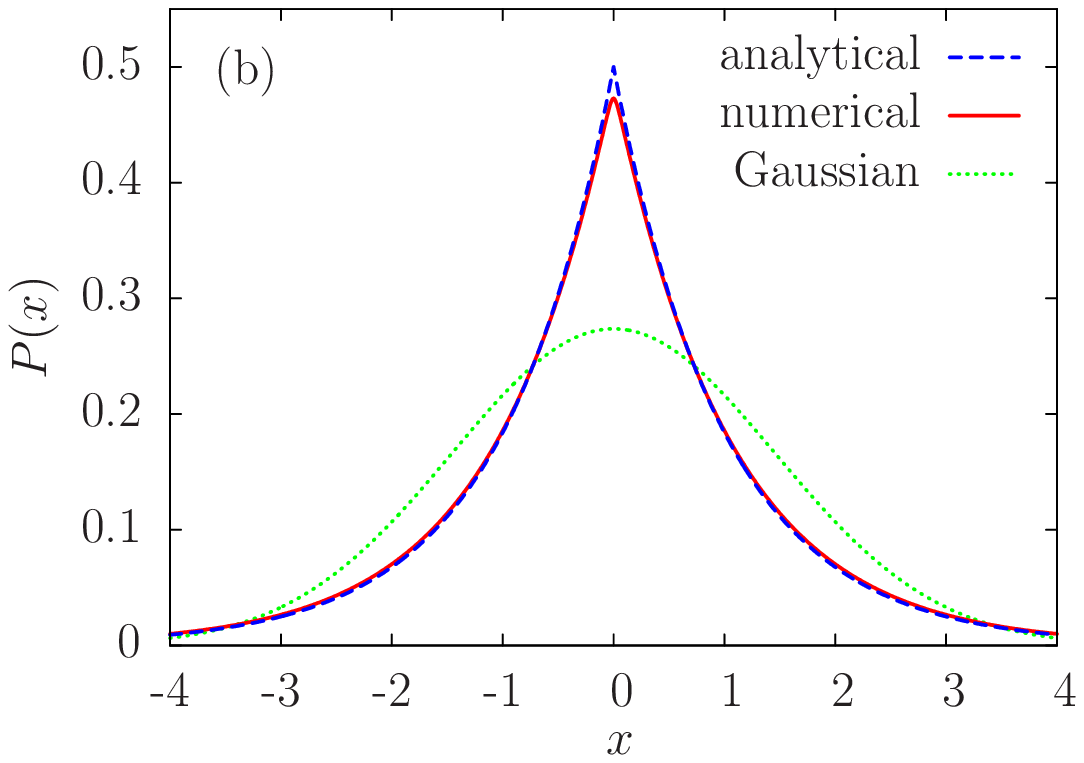}
\end{center}
\caption{\label{sub_super_density}  The probability distribution $P(x)$ at $t=1$
for two different viscosities:
In  (a)   $\eta(x)=0.01+x^2$  with
a minimum at $x=0$  and in (b)  $\eta(x)=(0.001+|x|)^{-1}$ 
with a maximum at $x=0$. The red line in each part is obtained by  numerical integration 
of (\ref{FPG}) and the nearly identical 
 blue dashed curve 
 is the analytical solution given by (\ref{FPanalytic})
for $\eta(x) \propto x^2$ in (a) and $\eta(x) \propto |x|^{-1}$ in (b).
In both parts $P(x)$ is
compared with a Gaussian profile (green dotted line) having the same second moment.
}
\end{figure}

There is a further  aspect to be drawn from the 
profiles of the probability distribution $P(x)$ in
the anomalous regime.
After the short normal diffusion in regime I,
$P(x)$ at $t=1$ deviates in regime II from a Gaussian profile
as shown in figure~\ref{sub_super_density}.
In part (a) the red curve of $P(x)$  is determined numerically 
by solving  (\ref{FPG})  for the spatially varying viscosity 
$\eta(x)=\eta_0+x^2$ with a small minimal viscosity $\eta_0=0.01$.
It is compared with
the analytical solution of the Fokker-Planck equation in
(\ref{FPanalytic}) for $\eta=x^2$ (blue dashed line) and we
find a nearly perfect agreement between both approaches 
in figure~\ref{sub_super_density}. Deviations  
grow by increasing the viscosity minimum $\eta_0$ in our simulations.
Both curves in figure~\ref{sub_super_density}(a)
deviate, however, 
significantly from a Gaussian distribution (green dotted line) with the same
second moment. 
As predicted in section \ref{FPE_analytic_solution} the distribution is platykurtic for $\beta>0$,
which can be explained as follows.
Due to the small minimal viscosity $\eta_0$
at the starting point at $x=0$, particles quickly diffuse away. This 
reduces the probability to find a particle at $x=0$ and simultaneously
enhances $P(x)$ in a wider neighborhood of the minimum of $\gamma(x)$, 
compared to the case of a constant damping with a Gaussian distribution.
 On the other hand, a viscosity increasing strongly  
as a function of $|x|$ impedes quick particle diffusion away 
from the minimum.
This reduces the probability  to find a particle at larger distances
compared to a Gaussian distribution.

With a  viscosity maximum at the starting point  of the
particles the situation is similar, as illustrated by the distribution 
in  figure~\ref{sub_super_density}(b).
Here the numerical solution $P(x)$ of the Fokker-Planck equation (red curve)
 is obtained
for the viscosity $\eta(x)=(0.001+|x|)^{-1}$ and  is compared with the
analytical solution given by (\ref{FPanalytic}) for $\eta(x) \propto |x|^{-1}$ (blue dashed curve). 
Only in the vicinity of the viscosity maximum at $x=0$
both approaches differ slightly. Again we find clear deviations
from a Gaussian distribution with an identical second moment 
as described by the green dotted curve in figure~\ref{sub_super_density}(b). 
Since the Brownian dynamics 
is significantly reduced near the maximum of the damping at $x=0$, the 
 particles move much slower away from the peak. This causes an enhancement
of $P(x)$ around the maximum, compared to a Gaussian distribution.
Thus the distribution is leptokurtic, in accordance with our predictions
in section \ref{FPE_analytic_solution}.

\begin{figure}[htb]
  \begin{center}
\includegraphics[width=0.49\columnwidth,angle=-0]{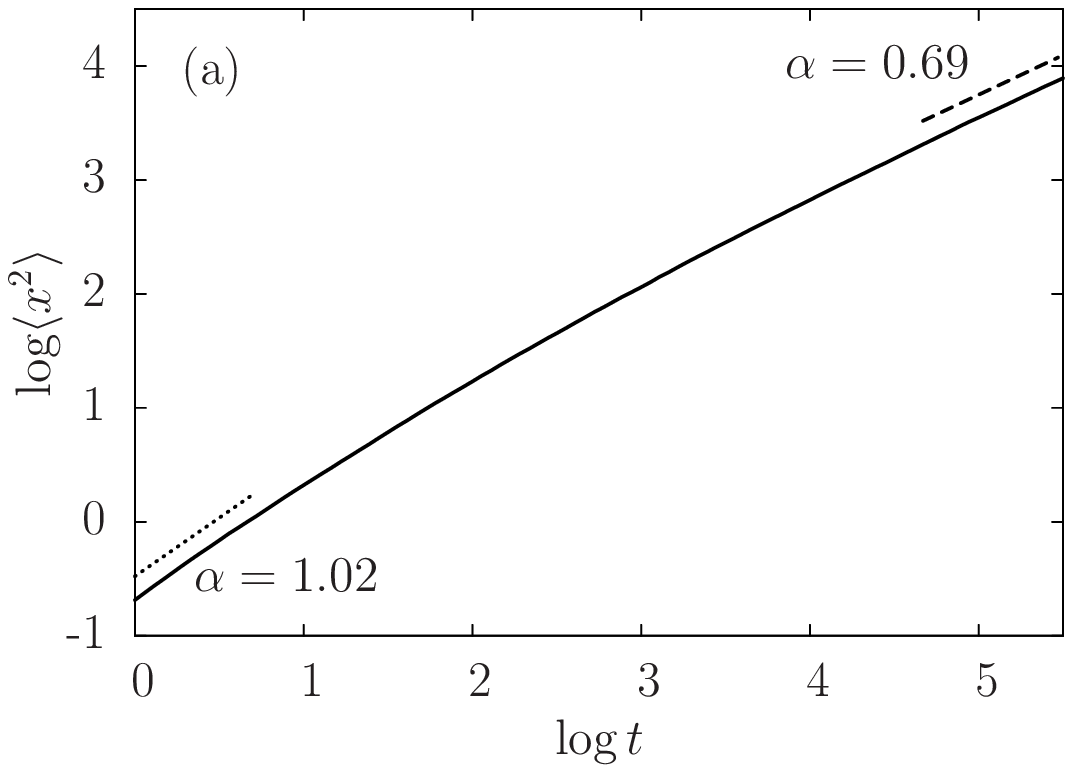}
\includegraphics[width=0.49\columnwidth,angle=-0]{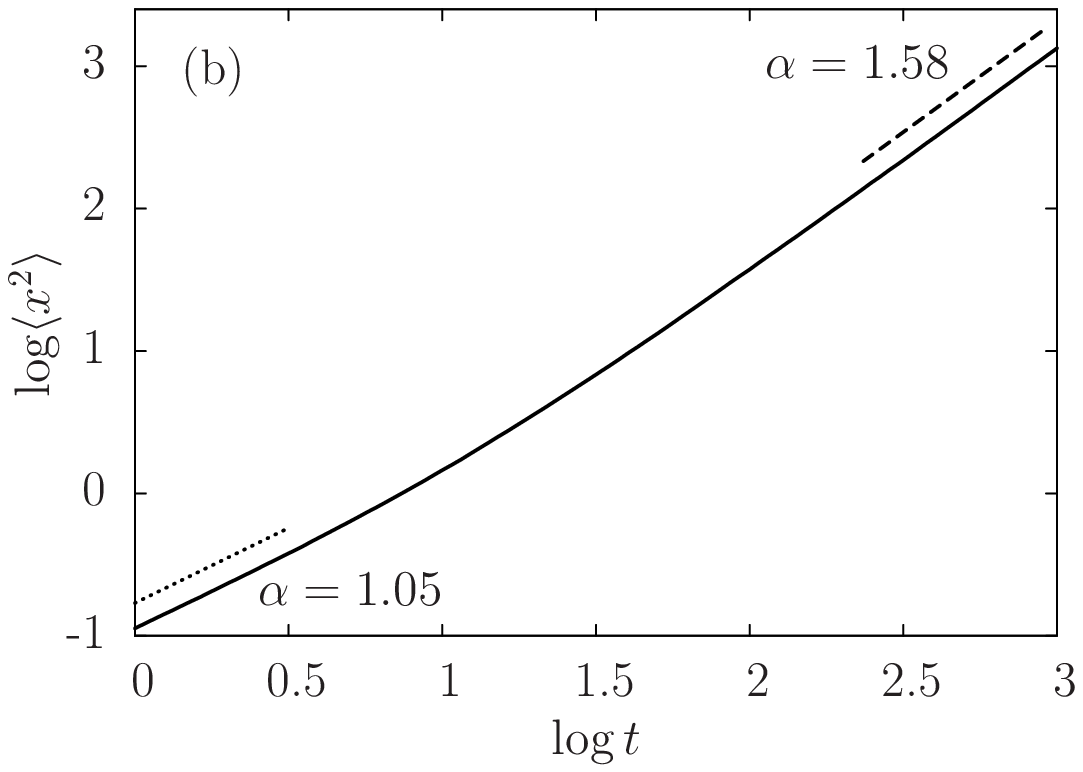}
  \end{center}
  \caption{\label{timescale} 
  The mean square displacement $\langle x^2(t) \rangle$
as obtained by  a numerical integration of (\ref{basicequation}) 
for two different model viscosities $\eta(x)$.
In (a) $ \eta(x) =10+|x|$ has a minimum at $x=0$, whereas 
in (b) one has for $|x|<1$ the plateau $\eta(x) =20$, which
is continued for $|x|\geq 1$
 by the decaying  function  $\eta(x)=20 |x|^{-3/4}$.
The dashed and dotted lines in both parts 
indicate linear fits to the numerical data,
whereby the slope $\alpha$ is explicitly shown.
}
\end{figure}

For a further analysis of the Brownian motion 
 the mean square displacement $\langle x^2(t) \rangle$ of the particle positions
is calculated by integrating the Langevin equation (\ref{basicequation})
numerically for an ensemble of particles starting at $x=0$. 
Figure~\ref{timescale} exemplarily shows $\langle x^2(t) \rangle $ 
as a function of time for two different spatially varying viscosities.
In each case and independent from the particular viscosity landscape,
the mean square displacement is characterized by the initial regime I of normal
diffusion, marked by the dotted lines.
Beyond this regime I, anomalous diffusion occurs
with exponents $\alpha \not = 1$, as long as the viscosity,
experienced by the particles,
 does not reach a non-vanishing constant plateau value,
as for instance for large $|x|$ in (\ref{viscb}).
In the vicinity of a viscosity minimum, subdiffusive Brownian
motion with exponents $\alpha < 1$ is found, as exemplified 
in figure \ref{timescale}(a) for $\eta(x)=\eta_0+\tilde \eta |x|$
with  $\eta_0=10$ and $\tilde \eta =1$. 
At long times the mean square displacement $\langle x^2 \rangle$ 
scales in the range of the dashed line
with $t^\alpha$ and $\alpha=0.69$, in good agreement with the
scaling result 
$\alpha=2/3$ predicted by formula  (\ref{msd_anomal}) with $\beta=1$.  
The viscosity used in figure~\ref{timescale}(b) has in 
a small range $|x|<1$  a constant maximum, $\eta(x)=20$, and 
then decays in the range $|x|>1$ 
according to the power law $\eta(x)= 20 |x|^{-3/4}$.
A fit to the numerical data in the limit of long times yields in the range
of the dashed line 
in figure~\ref{timescale}(b) the exponent $\alpha=1.58$,
deviating only slightly
from the scaling prediction $\alpha=1.6$ obtained
from (\ref{msd_anomal}) for  $\beta=-0.75$. 

The deviations between the
exponent obtained in simulations and the exponent obtained 
via scaling arguments decrease at sufficiently long times either
by reducing the minimal viscosity $\eta_0$ as in figure~\ref{timescale}(a)  or by shortening
the plateau in figure~\ref{timescale}(b).
The scaling regimes I (normal diffusion) and II (anomalous diffusion) 
 are separated by a transition period, where the crossover 
from $\alpha\simeq 1$ to $\alpha\neq 1$ takes place. 
%The location of the transition periods dependence of the boundaries
%of the anomalous regimes on the parameters has been estimated already by 
%  (\ref{charact_times_1}) and (\ref{charact_times_2}), respectively. 
This transition period may extend over several decades in
time as for instance in  
 figure~\ref{timescale}(a) with $ \tilde\eta / \eta_0 = 0.1$. Together with an
exponent $\beta=1$ this 
causes a transition period lasting over more than three decades.
In contrast, steep, nonlinear viscosity gradients obtained for $\beta>1$ and $\beta<0$ delimit the intermediate regime [see figure~\ref{timescale}(b)].

\begin{table}[htb]
%\begin{intended}
% \item[]
\begin{minipage}[tl]{0.45\textwidth}
\footnotesize
\begin{tabular}{@{}lccc}
\br
viscosity $\eta(x)$~~~~ &~~~$\beta$~ &  $\alpha$ (scaling) &  $\alpha$ (numerically) \\
\mr
$0.01+ |x|^4$      			& 4 	& 1/3 & 0.333 \\
$0.01 + |x|^2$     			& 2  	& 1/2 & 0.501 \\
$0.01 + |x|$       			& 1 	& 2/3 & 0.666 \\
$0.01 + |x|^{1/2}$ 			& 1/2 	& 4/5 	& 0.799 \\
$0.01 + |x|^{1/4}$ 			& 1/4 	& 8/9 	& 0.886 \\
\mr
$(0.01 + |x|^{1/2})^{-1}$ 		& -1/2 	& 4/3 	& 1.330 \\
$(0.01 + |x|^{2/3})^{-1}$ 		& -2/3	& 3/2 	& 1.494	\\
$(0.01 + |x|)^{-1}$ 			& -1 	& 2	& 1.976	\\
$(0.01 + |x|^{4/3})^{-1}$ 		& -4/3	& 3 	& 2.963	\\
\br
\label{tab:1}
\end{tabular}
\end{minipage}
%\end{intended}
\begin{minipage}[tr]{0.55\textwidth}
\caption{Chart of anomalous exponents $\alpha$
for spatially  varying viscosities $\eta(x)$ obtained
by the scaling formula (\ref{msd_anomal}) 
and via  numerical solutions of (\ref{basicequation}) by fitting
the resulting mean square displacement 
in regime II by the power law $\propto t^{\alpha}$.
We find a good agreement in both generic cases:
 For subdiffusion ($\alpha < 1$) with a minimum
of $ \eta(x)$  and for superdiffusion ($\alpha > 1$) 
with a maximum of $\eta(x)$.
}
\end{minipage}

\end{table}

For a further examination of the scaling prediction in the anomalous 
regime II, different spatially varying
viscosities were investigated  
numerically and in terms of the analytical 
scaling argument given by (\ref{msd_anomal}).
As can be seen in table~1, the numerically 
and analytically obtained exponents $\alpha$ 
are in excellent agreement for several  
 viscosities. In general, in the neighborhood of a 
viscosity minimum one obtains subdiffusive behavior, 
whereas in the vicinity of a viscosity maximum superdiffusive behavior is observed.
For the viscosities in table~1 with a small value $\eta_0$ at the
minimum or  a large value at the viscosity maximum one finds
according to  (\ref{charact_times_1}) and (\ref{charact_times_2})
an early onset of the intermediate anomalous regime II. 
We would like to emphasize, that the viscosities discussed up to now 
are approximations of (\ref{viscb}), which we used to demonstrate 
the good agreement between our scaling results and numerical 
simulations in the anomalous regime II. 
For the viscosity (\ref{viscb}) regime II is also limited in
time from above as described in the previous section by 
the characteristic times given in (\ref{charact_times_1}) and (\ref{charact_times_2}).

\begin{figure}[ht]
  \begin{center}
     \includegraphics[width=0.49\columnwidth,angle=-0]{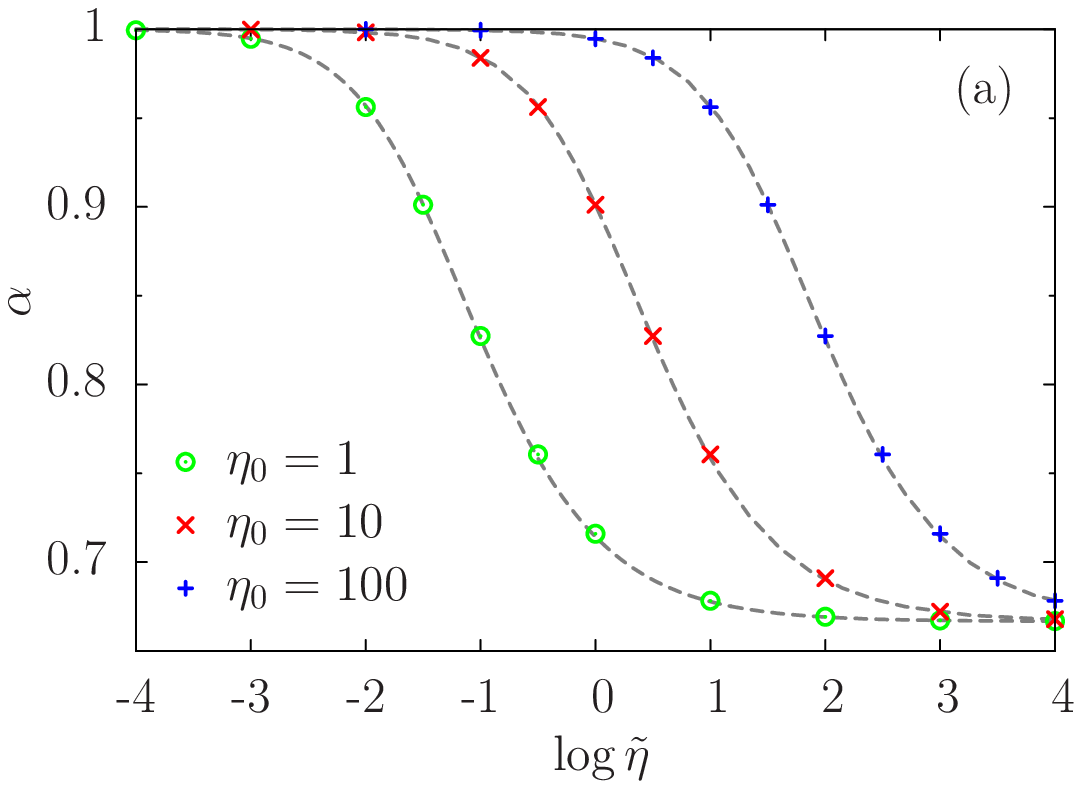}
     \includegraphics[width=0.49\columnwidth,angle=-0]{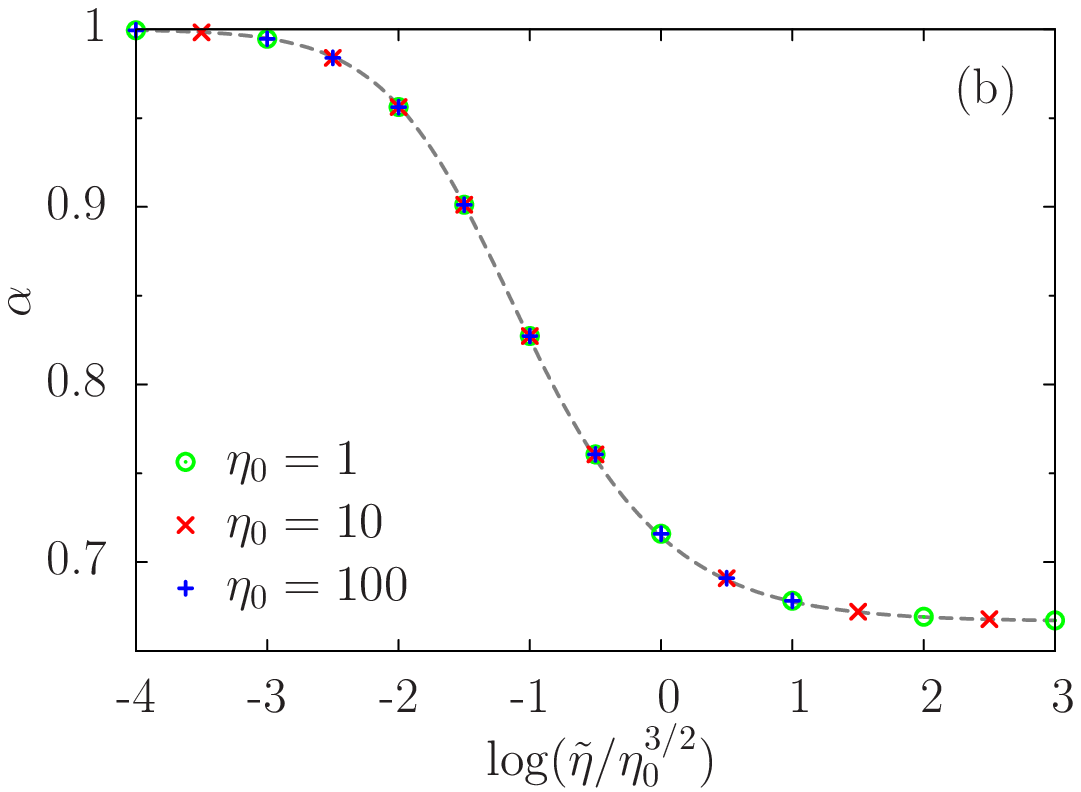}
  \end{center}
\vspace{-4mm}
  \caption{\label{exp1}
The scaling exponent $\alpha$ 
of the mean square displacement, 
$\langle x^2 \rangle \propto t^\alpha$, 
 is plotted in (a)
for the symmetric viscosity $\eta(x)=\eta_0 + \tilde \eta |x|$
 as a function of the slope $\tilde \eta$ for 
three different values of $\eta_0=1,~10,~100$ 
 and in part (b) as a function
of the ratio $\tilde \eta/\eta_0^{3/2}$. $\alpha$ is obtained by fitting
$\langle x^2 \rangle \propto t^\alpha$ over a fixed time interval $0<t<10^2$
to numerical data created by integration of equation (\ref{FPG}).
}
\end{figure}

To investigate the crossover  behavior between regime I and II,
the Langevin and the Fokker-Planck equation were
solved for the viscosity $\eta=\eta_0 +\tilde
\eta |x|$ up to the fixed time $T=100$ and for different values
of $\eta_0$ and $\tilde \eta$. With this approach we mimic
the limited time range available in experiments for detecting 
the anomalous regime. The  numerical curves of the
mean square displacement $\langle x^2(t) \rangle$ are fitted
over the whole range $0\le t \le T$ 
by the power law $t^\alpha$ and the resulting exponent
$\alpha$ is shown in figure~\ref{exp1} as function
of the viscosity. The deviation of $\alpha$
from $\alpha=1$ grows, when the anomalous  diffusion
regime II occupies an increasing part
 of the interval $[0,T]$, i. e. when the inequality  $\tilde \eta |x| > \eta_0$ holds
for an increasing part of $[0,T]$. 
The two terms $\tilde \eta |x|$ and  $ \eta_0$ in the denominator of (\ref{msq_approx}) become nearly equal 
at the time  $  t_t \sim \eta_0^3/\tilde \eta^2$. 
Hence, for decreasing values of $t_t$ the range
of anomalous diffusion in $[0,T]$ increases
as illustrated by the trend in figure~\ref{exp1}(a).  
On the other hand, the anomalous fraction
and therefore $\alpha$ can be kept constant by keeping
$\tilde\eta^2 /\eta_0^3$ constant, which is shown by
 figure~\ref{exp1}(b).

The behavior of $\alpha$
 in figure~\ref{exp1}  can also be derived via (\ref{msq_approx}) by
 using $\eta=\eta_0+\tilde \eta \langle x^2 \rangle^{1/2}$.
The resulting nonlinear equation for the second moment,
$ \langle x^2\rangle ( \eta_0 + \tilde\eta \langle x^2 \rangle^{1/2})= Q t$, 
can easily be solved. From this solution the exponent $\tilde \alpha(t)$, which
describes the local slope along curves as  in figure~\ref{timescale},
can be calculated by  $\tilde \alpha(t) = d \log(\langle x^2 \rangle)/ d \log(t)$. 
Its average $ \alpha = \int_0^T dt ~\tilde \alpha(t)$ corresponds to
the dashed lines in figure~\ref{exp1}, which agree
 surprisingly well with the full numerical results obtained by (\ref{FPG}).
 We find visible deviations from the full solution only if the viscosity shows
very small values close to zero either at the minimum or for $|x| \gg L$, which, however,
is unlikely in experiments. Therefore,
  the exponent obtained via the scaling 
relation (\ref{msq_approx}) may be useful in analyzing
and fitting  experimental results.

\begin{figure} [htb]
  \begin{center}
     \includegraphics[width=0.49\columnwidth,angle=-0]{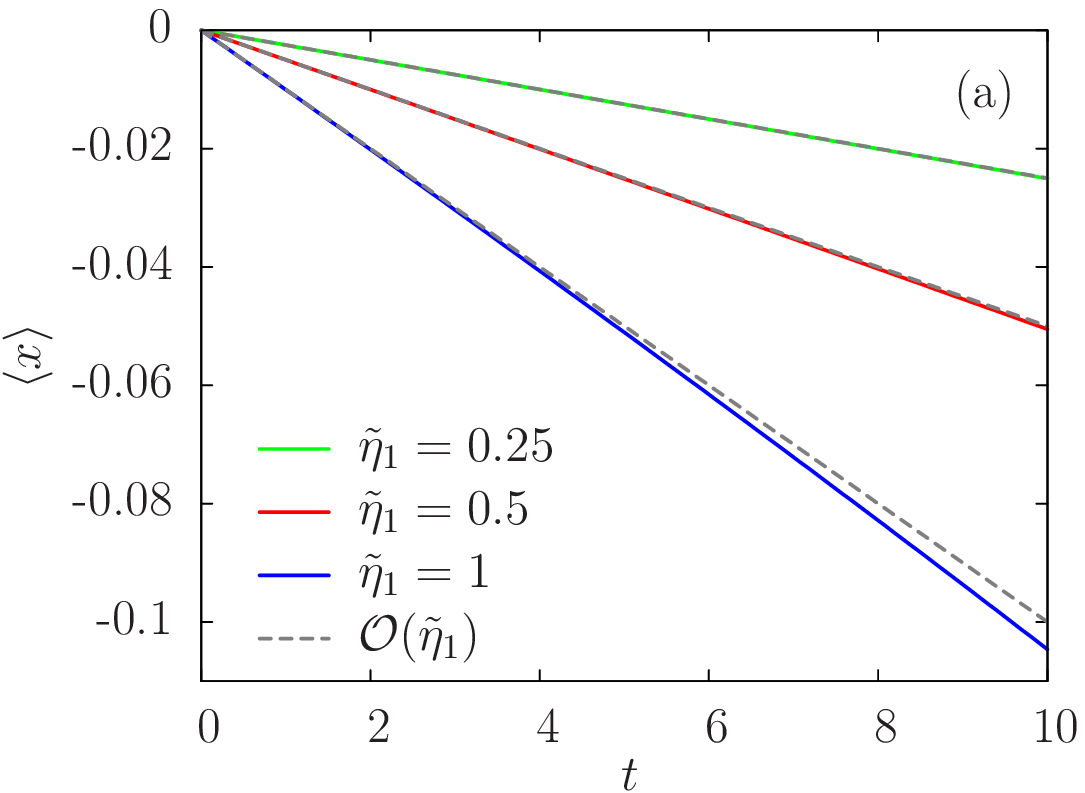}
     \includegraphics[width=0.49\columnwidth,angle=-0]{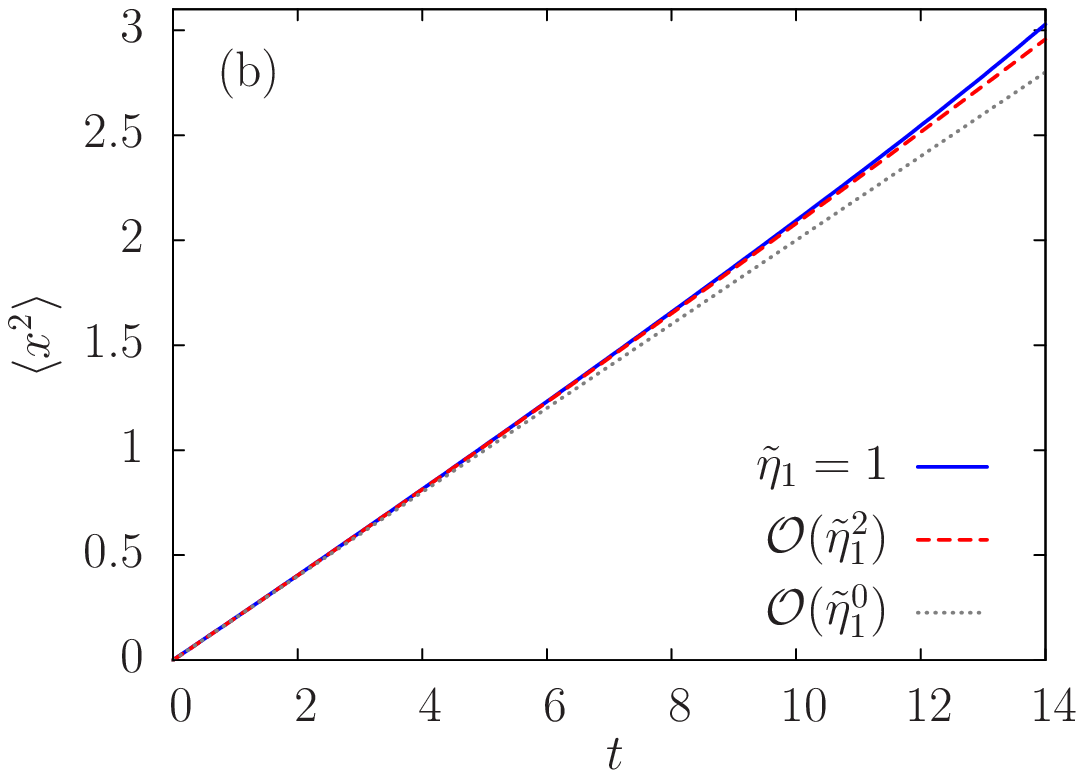}
  \end{center}
\caption {\label{anadrift} 
The first two moments of the numerical solutions of (\ref{FPG})  (solid lines), $\langle x \rangle$ in part (a) 
 and $\langle x^2 \rangle$  in (b),
are compared with the leading order contributions to the approximations given by 
 (\ref{mean_asym})  respectively (\ref{msd_asym}) (dashed lines) 
for a linear viscosity profile $\eta(x)= \eta_0 + \tilde \eta_1 x$ with $\eta_0=10$ 
and different values of $\tilde\eta_1=0.25,~0.5,~1$.
}
\end{figure}

A further experimental  relevant model 
viscosity is described by the asymmetric function (\ref{visctanh}), which
may be approximated in the range $|x/L| \ll 1$ by the linear dependence in
equation (\ref{visca}).
 The essential
qualitative difference to the previous model viscosities is its
asymmetry with respect to the starting point of the particles 
at $x=0$, which causes 
 non-vanishing moments $\langle x \rangle$ and 
$\langle x^3 \rangle$ as given in 
(\ref{mean_asym}) and (\ref{third_asym}) for short time intervals.
The quality of  the approximate solutions in (\ref{mean_asym})-(\ref{third_asym})
is investigated  in  figure~\ref{anadrift}, where
in part (a) 
the leading contribution
to the drift of the mean value in  (\ref{mean_asym}) (dashed lines) is compared 
 to the first moment 
of the numerical solution of (\ref{FPG}) 
(solid lines). Despite the finding that the differences between both approaches increase
 with $\tilde \eta_1 t$, the results are still in good
agreement since the deviations do not exceed  $0.25 \%$ for $\tilde \eta_1 =0.25$
and  $4.5 \%$ for $\tilde \eta_1=1$, even at $t=10$. 

The diffusion in an asymmetric viscosity profile (\ref{visca}) becomes
  anomalous too, as indicated  
by the contributions $ \propto t^2$ and $\propto t^3$ 
to the approximate second  moment in equation (\ref{msd_asym}). 
Again,  the deviations are small between the
full numerical result of $\langle x^2(t)  \rangle $ 
[solid line in figure \ref{anadrift}(b)] and 
the approximation (\ref{msd_asym}) [dashed  line in figure \ref{anadrift}(b)].
 We point out, that anomalous diffusion still persists, when the drift of
 $\langle x \rangle$ is subtracted from the
particle dynamics as illustrated by 
the following formula:   
$ \langle x^2 \rangle - \langle x \rangle^2 
= \frac{2\tilde D}{\eta_0}~ t + \frac{\tilde\eta_1^2 \tilde D^2}{\eta_0^4} ~t^2
+ {\cal O}(\tilde \eta_1^4)$.  
With the help of the third moment 
(\ref{third_asym}) one can also define the skewness $S$ of a 
probability distribution \cite{SokolovRad:2009}, which is given in our
case at the leading order 
 by  $S=-(3\tilde \eta_1 \sqrt{\tilde D t})/(\sqrt{2} \eta_0^{3/2} )$. 
Since $S$ is negative
for $\tilde \eta_1>0$, the left tail of the probability distribution becomes more important.

\begin{figure}
  \begin{center}
     \includegraphics[width=0.49\columnwidth,angle=-0]{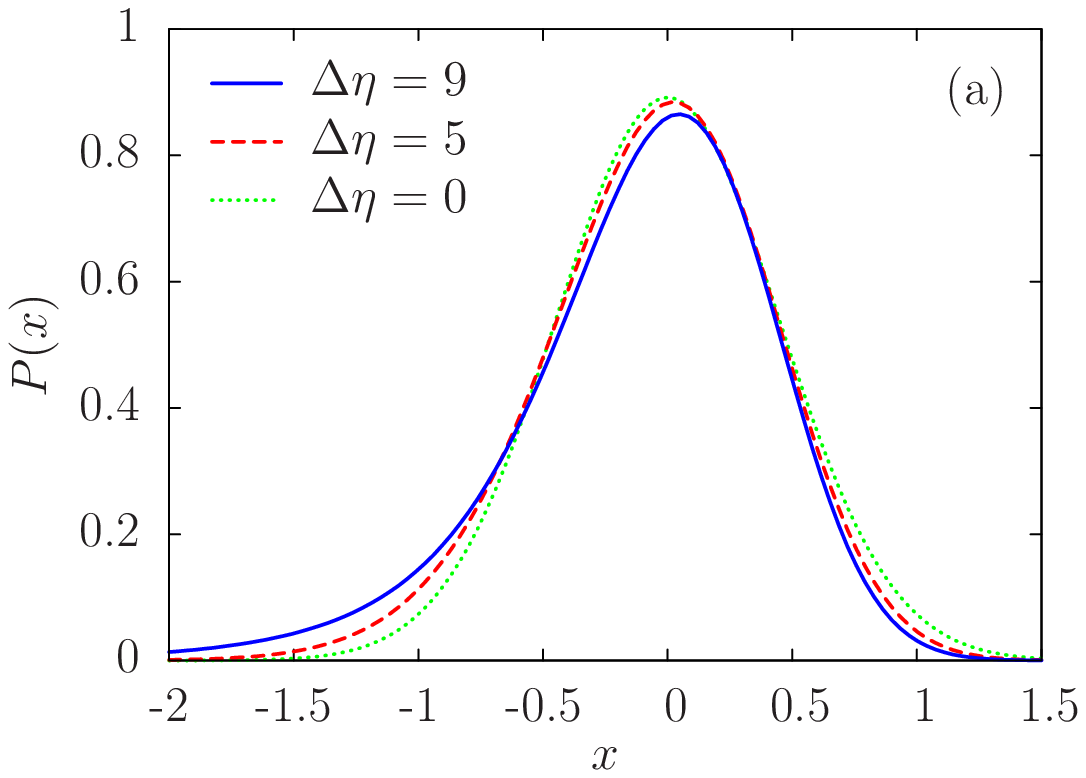}
     \includegraphics[width=0.49\columnwidth,angle=-0]{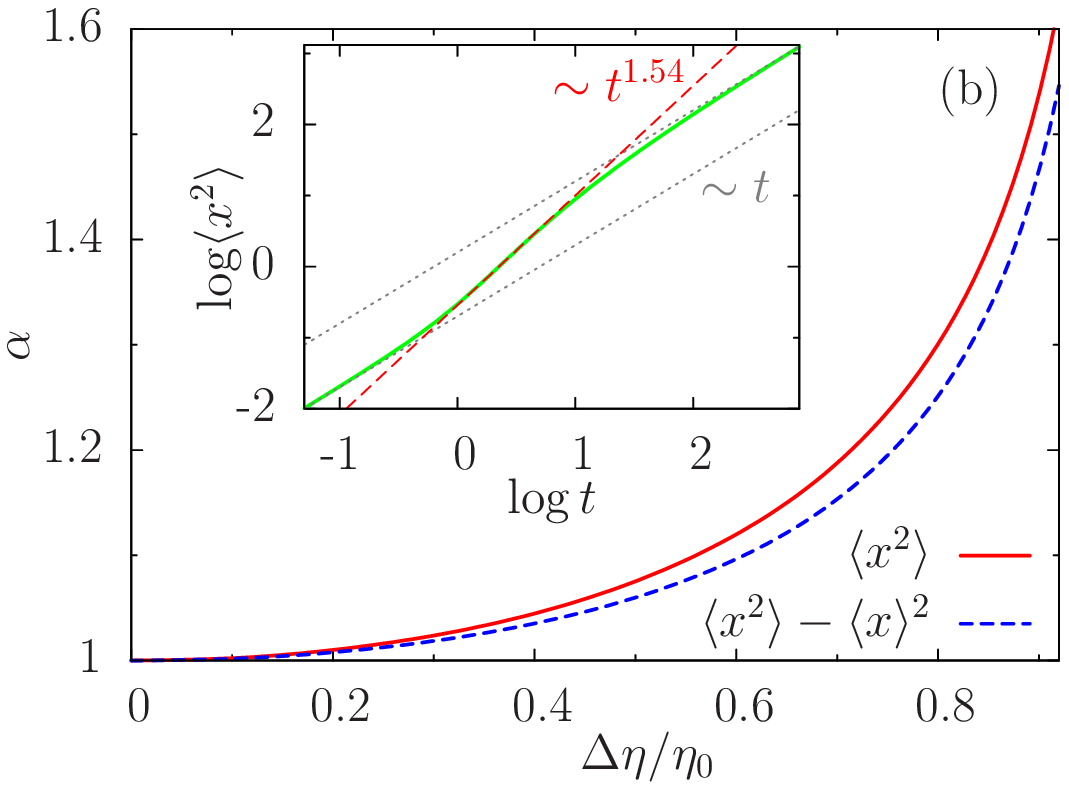}
  \end{center}
\caption{  \label{Flinvis}
The probability distribution $P(x)$ at $t=1$
in (a) and the mean square displacement $\langle x^2 \rangle$ in (b)
are  obtained by integrating (\ref{FPG})  for the 
asymmetric viscosity  (\ref{visctanh}) 
with  $\eta_0=10$ and different values of $\Delta\eta=0,~5,~9$. 
The inset in (b) shows the three regimes of $\langle x^2(t) \rangle $ 
for $\eta_0=10$ and  $\Delta\eta=9$.  
In the intermediate regime II  at the largest slope (dashed line)
the scaling exponent $\alpha$ is determined by the fit 
$\langle x^2 \rangle \propto t^\alpha$ to the  numerical data.
In (b) the resulting $\alpha$ in regime II is given as function
of $\Delta \eta/\eta_0$:
The red solid line belongs to the evolution of $\langle x^2 \rangle$
and the dashed line to the variance $\langle x^2\rangle - \langle x \rangle^2$.
}
\end{figure}

All three aspects, the non-vanishing drift, the skewness of the particle
distribution and the anomalous diffusion caused by an asymmetric monotonous
viscosity profile are  recaptured  in numerical solutions of the Langevin equation  (\ref{basicequation})
and the Fokker-Planck equation (\ref{FPG}) by  integrating both
for the viscosity  (\ref{visctanh})
up to time scales beyond the validity range of equations (\ref{mean_asym})-(\ref{third_asym}).
The asymmetry of the resulting probability distribution is shown 
in figure~\ref{Flinvis}(a) for
increasing values of $\Delta \eta$,
where the left tail is the more important one as predicted by the 
skewness $S$ of the distribution introduced above.
The time dependence of the mean square displacement $\langle x^2(t) \rangle$ 
is shown for the viscosity (\ref{visctanh}) by the inset in
figure \ref{Flinvis}(b) for $L=1$,  $\eta_0=10$ and $\Delta \eta=9$ and
illustrates  all three temporal diffusion regimes. During the early temporal
regime I with $|x|\ll L$ and in the long time limit with $|x|\gg L$  (regime III)
one has normal diffusion with the exponent $\alpha=1$
as indicated by the dotted lines in the inset of  \ref{Flinvis}(b).
In contrast, during the intermediate regime II with $|x| \simeq L$ one has superdiffusion
with $\alpha =1.543 >1$ as indicated by the dashed line. 
If the numerical data of the mean square displacement, obtained 
for $\eta_0=10$ and $\Delta \eta=9$,
are fitted  by $\langle x^2 \rangle \propto t^\alpha$ 
in the range of its largest slope in regime II we find that
$\alpha$ is independent of the length $L$ as indicated
by the values $\alpha=1.538 ~(L=0.2)$, $\alpha=1.543~(L=1)$, and 
$\alpha=1.544~(L=5)$.  
However, the exponent $\alpha$ varies
strongly as a function $\Delta \eta/\eta_0$ in figure \ref{Flinvis}(b). 
As predicted by the analytical
expressions in (\ref{mean_asym}) 
and (\ref{msd_asym})
 the mean square displacement $\langle x^2 \rangle$ [see also figure \ref{anadrift}(b)]
and the variance $\langle x^2 \rangle - \langle x \rangle^2$  show
anomalous diffusion behavior as well.  The solid line in figure \ref{Flinvis}(b)
describes the exponent $\alpha$ of the mean square displacement 
and the lower lying dashed line is the exponent related to the variance
 $\langle x^2 \rangle - \langle x \rangle^2$. For both quantities the
 exponent $\alpha$
 may take even larger values  in the limit $\Delta \eta /\eta_0 \to 1$. 
The mean value  $\langle x \rangle$ drifts also for the
viscosity profile  (\ref{visctanh}). 
For small values of $t$ and $x$  we find a perfect agreement
with the linear time dependence given by the leading contribution in
 (\ref{msd_asym}), but its dependence becomes more complex in the
range of larger values of $\Delta \eta$ and in  the long-time limit.

%%%%%%%%%%%%%%%%%%%%%%%%%%%%%%%%%%%%%%%%%%%%%
\section{Conclusions}
\label{conclusion}
%%%%%%%%%%%%%%%%%%%%%%%%%%%%%%%%%%%%%%%%%%

We have identified three different 
diffusion regimes
for  mesoscopic Brownian particles in spatially varying viscosities,
which were  analyzed by three approaches: 
Firstly, by scaling arguments applied 
to the expression of the mean square displacement, 
secondly by  simulations of the corresponding nonlinear 
Langevin equation~(\ref{basicequation}), 
and thirdly by solving the related Fokker-Planck equation~(\ref{FPG})  
either numerically or, in limiting cases, even analytically. 
For an ensemble of particles starting at a viscosity extremum a short regime
of normal diffusion is found where the mean square displacement $\langle x^2 \rangle \propto t^\alpha$ scales  with the exponent $\alpha=1$.  
Beyond this regime   Brownian 
particles experience considerable changes of the viscosity
along their trajectories, which leads to anomalous 
diffusive motion in an intermediate temporal regime. 
Near a minimum of the viscosity 
the particle dynamics becomes subdiffusive  
with an exponent $\alpha<1$, whereas it becomes superdiffusive with
$\alpha > 1$ in the vicinity of viscosity maxima.
In the long time  limit, when the Brownian particles explore
the whole range of a viscosity variation, as for instance described by
(\ref{viscb}) or for a periodically varying viscosity, the
particles experience a mean  
viscosity and therefore normal diffusion is found. 

In the intermediate anomalous diffusive regime
the particles' probability distribution is non-Gaussian.
In the case of a viscosity minimum the particle distribution 
function is - compared to a Gaussian with the same second moment -
reduced at its maximum, increased at intermediate distances from
its maximum, and reduced again at large distances.
The opposite happens near a maximum of the viscosity: 
The particle distribution function is enhanced at
the starting point of the particles and reduced at intermediate distances 
from the initial position. 
For an ensemble of particles starting in the range of a linearly
varying viscosity profile, which is asymmetric with respect to
the starting point,  one finds 
an asymmetric particle distribution
reflecting itself in a non-vanishing skewness.
In addition, the first moment of the distribution
drifts as function of time into the direction of decreasing
viscosity and the particles show 
superdiffusive behavior. However, superdiffusivity is only partly caused
by the particle drift: The exponent of the
superdiffusive motion reduces only slightly towards 1 when the mean drift
is subtracted from the particle motion but still shows clearly
superdiffusive behavior with $\alpha>1$.

The subdiffusive behavior in an intermediate regime
 in the neighborhood of a minimum of
the particle  damping shares  similarities with
the Brownian dynamics of a single segment of a flexible polymer,
the so-called Rouse dynamics.  Also in this case one has
normal diffusion of the segments on a short as well as on a long time scale. 
In the intermediate range,
where the mean square displacement of the segment
is of the order of the polymer-coil diameter, one
finds the subdiffusive behavior
 $\langle x^2 \rangle \propto t^{1/2}$ \cite{deGennes:79}.

In contrast to most of the well-known examples showing 
anomalous diffusion, one can imagine for systems suggested in this work
both types of anomalous diffusion, 
sub- and superdiffusion.
In photorheological materials 
\cite{Raghavan:2007.1}, for example, 
either a maximum or a minimum of the viscosity as well as monotonously varying
viscosities can be induced by an appropriate 
spatial variation of the illumination strength.
In some binary mixtures the sign of the Soret effect changes
as a function of the mean temperature of the mixture \cite{Wiegand:2004.1}. 
The two possible signs may be used to attract either 
the lower viscous component to the locally heated area or
 the higher viscous component.
In glass forming polymer-mixtures the relation between
the local composition and the viscosity
can be  a nonlinear function  \cite{Williams:55.1,KoehlerW:2002.1}. 
Materials with strong variations of the local viscosity are especially
 favorable to observe anomalous diffusion, as discussed in this work.

Another interesting subject are heated particles in a binary fluid
mixture, undergoing Brownian 	
motion. An example are light absorbing particles 
in a transparent binary fluid-mixture with an upper 
miscibility gap such as  for instance 
$n$-Butoxyethanol and water \cite{aizpiri:1992.1}, 
where the temperature may be driven
close to the transition temperature.
The inhomogeneous temperature field around a particle sets
in much faster than 
 related changes of the concentration and the viscosity  in the neighborhood
of the particle,
% . In addition the spatial dependence of concentration and viscosity
%  is  expected to vary on a
% considerably larger length scale than the temperature, so
such that the Brownian particle experiences temporally its 
own induced viscosity variations. The delayed dynamics between 
temperature and viscosity field may lead to interesting
memory effects that are also prone to cause anomalous diffusion
% as
and will be investigated in forthcoming work.

%%%%%%%%%%%%%%%%%%%%%%%%%%%%%%%%%%%%%%%%%%%%%
\section*{Acknowledgments}
%%%%%%%%%%%%%%%%%%%%%%%%%%%%%%%%%%%%%%%%%%%%%

This work was started during a summer project for undergraduate students and
was supported by the German Science Foundation via 
the research unit FOR608, the research center SFB 481 and 
the priority program on micro- and nanofluidics SPP 1164.
We thank J. Bammert, D. Kienle, W. Pesch, and S. Schreiber for useful discussions.

\vspace{6mm}

\bibliographystyle{prsty}

\end{document}